\documentclass[12pt,a4paper]{article}
\usepackage{amsmath}
\usepackage{graphicx}
\usepackage{times}
\begin{document}
\begin{titlepage}
\title{Are there  any clues for hadron asymptotics in the LHC data?}
\author{ S.M. Troshin, N.E. Tyurin\\[1ex]
\small  \it SRC IHEP of NRC ``Kurchatov Institute''\\
\small  \it Protvino, 142281, Russian Federation}
\normalsize
\date{}
\maketitle

\begin{abstract}
We  discuss  several clues related to the possible asymptotic behavior of hadron interactions arising on the base of performed analysis of the recent TOTEM data at the LHC and emphasize that the deep-elastic scattering can provide an important information on the asymptotical dependence via the unitarity relation. 
\end{abstract}
\end{titlepage}
\setcounter{page}{2}

The geometrical properties of hadron interactions are often attributed to the impact--parameter dependence of the inelastic overlap function.
 For  long period of time the elastic scattering  was consistent with the so called BEL picture where the protons' interaction region becomes blacker,  edgier and larger \cite{valin1}.   However, as it was noted on the model ground  \cite{edn}, the inelastic overlap function at the asymptotical energies could have a peripheral dependence on the impact parameter. This peripherality was interpreted as a manifestation of emerging interaction transparency in the central collisions. Later on, this interpretation has been generalized and specified in papers \cite{bd1,bd2,bd3} where this phenomenon  was correlated with antishadowing or reflective scattering in hadron interactions.  A recent analysis \cite{alkin} of the elastic scattering data obtained by the TOTEM  at $\sqrt{s}=7$ TeV has revealed an existence  of this novel feature in strong interaction dynamics due  to transition  at such high energies to a new scattering mode  \cite{bd1,bd2,bd3,degr,anis} referred nowadays as antishadowing, reflective or resonant scattering.

A slow gradual transition to the emerging REL picture, i.e. picture where the interaction region starts to become reflective at the center ($b=0$) and simultaneously becams relatively edgier, larger and black at its periphery, seems to be observed by the TOTEM experiment under the measurements of the $d\sigma/dt$ in elastic $pp$--scattering. There are several phenomenological models able to reproduce such transition and among them the model  based on the rational unitarization of the leading vacuum Regge--pole contribution with intercept greater than unity, described in 
\cite{edn}, and similar models   labelled by the generic name of the unitarized supercritical Pomeron (cf. \cite{anis} for a recent  discussion and the references).

There are two main questions one can pose now. Which limit for the scattering amplitude is saturated, the black disk limit or the unitarity limit. The second question is based on assumption of the unitarity limit saturation, namely, what is the value of the energy where the black limit for the scattering amplitude is crossed. The analysis performed in \cite{alkin} indicates that the black-disk limit has already been crossed at 
$\sqrt{s}=7$ TeV, while a simple extrapolation based on the supposed gaussian dependence of the elastic scattering amplitude on the impact parameter assumes a delayed crossing happening at higher energies \cite{drem}. Evidently, such extrapolation is at variance with large $-t$ data measured at $\sqrt{s}=7$ TeV. 

In this short note we are going to discuss the two above  questions.
 The first problem deals with possible saturations of two limits: black--disk limit or unitarity limit which is twice as large as a former one. It seems natural to suppose a monotonic increase of the elastic scattering amplitude $f(s,b)$ with the energy increase without any oscillations. The analysis \cite{alkin}
 provides a hint that the unitarity limit and not a black disk one is saturated in this case. This analysis has shown that
$f(s,b)$ becomes greater than the black-disk limit of $1/2$ at $\sqrt{s}=7$ TeV, but the relative excess $\alpha$ ($f(s,b)=1/2[1+\alpha(s,b)]$) being still rather small at this energy and small impact parameters. The value of $\alpha$ is only about $0.08$  at $b=0$ \cite{alkin}. 
Hence, the most relevant objects to study  an initial crossing of the black-disk limit are the functions $f(s,b)$ and $h_{el}(s,b)$, and not the inelastic overlap function $h_{inel}(s,b)$) since relative negative deviation at small values of $b$ in the latter case has an order of $\alpha^2$, i.e. $h_{inel}(s,b)=1/4[1-\alpha^2(s,b)]$, where $\alpha(s,b)$ being non--vanishing in the region $0\leq b<r(s)$. The functions $f$, $h_{el}$ and $h_{inel}$ enter the unitarity relation in the impact parameter representation. This relation has the following form:
\begin{equation}\label{un}
\mbox{Im}f(s,b)=h_{el}(s,b)+h_{inel}(s,b).
\end{equation}
Using common assumption on the pure imaginary scattering amplitude\footnote{It should be noted here that saturation of the black--disk limit or the unitarity limit leads to a vanishing real part of the scattering amplitude, $\mbox{Re} f\to 0$. This vanishing is valid in the two above limiting cases \cite{tr}.} (and replacing $f\to if$) at high energies, one obtain that 
\begin{equation}\label{un1}
h_{inel}(s,b)=f(s,b)[1-f(s,b)].
\end{equation} 
From this relation one can easily obtain that correction to this function is indeed of order of $\alpha^2$.

The analysis \cite{alkin} based on the consideration of the {\it differential} cross--sections is a most sensitive method of the impact picture reconstruction.  However, since the deviation $\alpha$ is small, the effective reassurance in  validity of conclusion on the black disk limit crossing requires other possible indications which would testify in favor of  the black-disk limit overshooting and the  subsequent unitarity limit saturation. This saturation corresponds to the limiting case  when $ S(s,b)\to -1$ at fixed $b$ and $s\to \infty$ and can be referred as a pure reflective scattering by analogy with the reflection of light in optics  \cite{bd3}. 
The appearance of the reflective scattering is associated with increasing central density of the colliding protons with the energy growth. It can be said that beyond some critical value of the density, which correspond to the black-disk limit, the colliding protons start acting like the hard billiard balls. Such behavior can be compared to a reflection of the incoming wave by a metal (with change of phase of the incoming wave by $180^0$ due to presence of free electrons).  An increasing reflection ability appearing in that way is to be associated with  a decreasing absorption according to the probability conservation expressed in the form of unitarity relation. 
The principal point of the reflective scattering  is  fulfillment of the inequalities  $1/2  < f(s,b) <1$ and $0  > S(s,b) > -1$, those allowed by the unitarity relation \cite{bd1,bd2}. Our goal here is to discuss some of the experimental manifestations of this novel scattering regime. 
 
 As it was already noted, exceeding the  black-disk limit is a principal conclusion of the model--independent analysis of the impact--parameter dependencies of the amplitude performed in \cite{alkin}.  
The amplitude  at small $b$ values is sensitive to the $t$--dependence of the scattering amplitude $F(s,t)$ in the region of large values of $-t$ (referred as deep--elastic scattering \cite{islam}).   As it was shown in \cite{deepel} the saturation of the unitarity limit leads to the relation
\begin{equation}\label{asmpt}
(d\sigma^{rfl}_{deepel}/dt)/(d\sigma^{abs}_{deepel}/dt)\simeq 4.
\end{equation}
However, at the LHC energy $\sqrt{s}=7$ TeV the difference is not so significant and the positive deviation from the black-disk limit is small 
\begin{equation}\label{asmpt}
(d\sigma^{rfl}_{deepel}/dt)/(d\sigma^{abs}_{deepel}/dt)\sim 1+2\alpha(s,b=0).
\end{equation}
The models based on absorptive approach which cannot reproduce crossing of the black-disk limit in, in particular, the eikonal models, provide a rather poor description of the LHC data in the deep--elastic scattering region (cf. e.g. \cite{b2,godiz,drufn}). 
However, without a correct description of the differential cross-section in this region of $-t$ the existing crossing of the black-disk limit can be easily missed.  This fact emphasizes also importance of consideration of the {\it unintegrated} quantities. In contrast,
the Donnachie-Landshoff  model \cite{dln}, where the black disk limit is exceeded, is in a good agreement with the new experimental data on 
$d\sigma/dt$ at $\sqrt{s}=7$ TeV in the whole region of transferred momentum. The same is true for the models based on the rational or $U$--matrix form of unitarization. Fig. 1 (borrowed from 
\begin{figure}[hbt]
\begin{center}
\hspace{0.5cm}
\resizebox{10cm}{!}{\includegraphics*{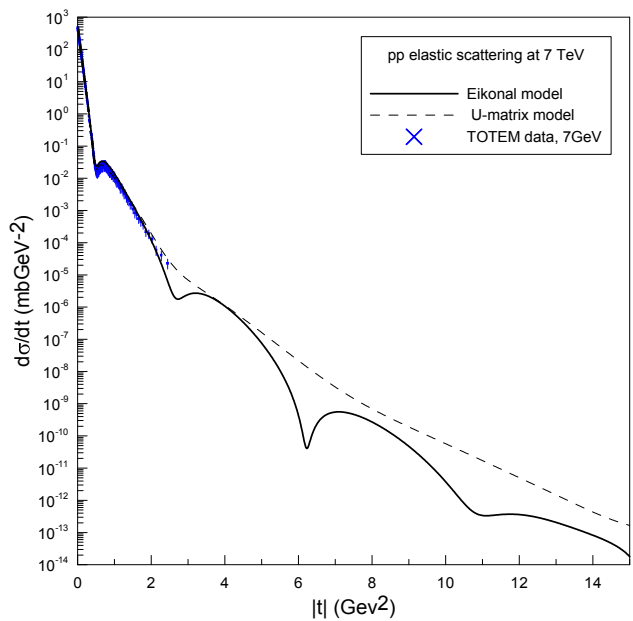}}
\end{center}
\vspace{-3.5cm}
\caption[ch1]{\small Description \cite{martyn} of differential cross--section of $pp$-scattering at $\sqrt{s}=7$ TeV in absorptive (solid line) and reflective (dashed line) forms of unitarization.}
\end{figure}
\cite{martyn}) is a kind of illustration of the above statements. As it is clear from Fig. 1, the absorptive (eikonal)  models (based on the exponential unitarization) predict lower values for the differential cross-section in $pp$-scattering at $\sqrt{s}=7$ TeV and appearance of the secondary bumps and dips  at large values of $-t$.

 The reasons for the above qualitative difference in the behavior of $d\sigma/dt$ in the deep-elastic scattering region become apparent from the consideration of the scattering amplitude in the impact parameter representation for the two forms of unitarization.
 \begin{figure}[hbt]
\begin{center}
\resizebox{12cm}{!}{\includegraphics*{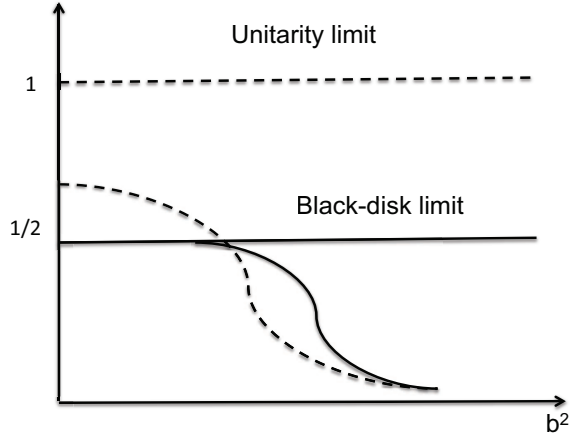}}
\end{center}
\vspace{-1cm}
\caption[ch1]{\small Schematic impact-parameter dependence of the amplitude $f(s,b)$ in the two cases of absorptive (solid line) and reflective (dashed line) forms of unitarization.}
\end{figure}
In the pure imaginary case (considered here) the following relation takes place
\begin{equation}\label{stot}
f(s,b)=\frac{1}{4\pi}\frac{d\sigma_{tot}}{db^2}.
\end{equation}
The following observations can be made:
\begin{itemize}
\item First,  the models based on absorptive form of unitarization  have an almost flat $b$-dependence in the region of the small and moderate impact parameters since at the energy of $\sqrt{s}=7$ TeV the black-disk limit is already reached at $b=0$  \cite{drufn} (cf. Fig.2). The result of such flatness is the sequence of the secondary dips and bumps (solid line, Fig.~1). Moreover,
the imaginary part of the amplitude $F(s,t)$ would give a dominant contribution in $d\sigma/dt$ at large values of 
$-t$.

\item Second, in any unitarized model (absorptive or reflective)  reproduction of the experimentally observed total cross-section requires that the areas under the solid and dashed curves (cf. Fig. 2) are to be equal. We believe that the dashed curve closely corresponds to the experimental situation. Then, to reproduce experimental value of the total cross-section the solid curve being almost flatten at small $b$ (it cannot exceed $1/2$)  should be shifted  to the larger values of the impact parameter as it is depicted on Fig. 2.  The slope parameter $B(s)$ \[B(s)\equiv \frac{d}{dt}\ln \frac{d\sigma}{dt}|_{-t=0}\]  is determined by the average value  $\langle b^2\rangle$. Therefore, one can conclude that the experimental value of this parameter cannot be well reproduced in this case.  It has already been mentioned that  excess above the black disk limit is not very significant at $\sqrt{s}=7$ TeV, but we expect its increase at higher energies. That means that the difficulties in the simultaneous description of the total cross-section and the slope parameter  in absorptive (eikonal) models would become more noticeable with the collision energy growth. Indeed, widening distribution over impact parameter space leads to the narrowing of the corresponding distribution over transferred momentum. 
\end{itemize}

Thus, the two discussed options---saturation of the black disk limit or saturation of the unitarity limit---generate different mechanisms of the total cross--section growth at the energies $\sqrt{s}>7$ TeV. Namely, saturation of the black disk limit assumes the growth of the total cross--section occurs due to increasing impact parameter values only, while the growth of total cross-section can, in fact, be due to both factors --- increase of impact parameter values combined with an increase of the elastic scattering amplitude $f(s,b)$ with energy at fixed $b$.   Therefore, the measurements of the elastic scattering at $\sqrt{s}=13$ TeV (especially in the region of large $-t$)  could be quite decisive 
regarding this aspect of asymptotic picture and would help to determine which  scattering picture ---absorptive or reflective --- should be expected at the asymptotical energies.
\section*{Acknowledgements}
We are grateful to E. Martynov for possibility to use the results  of ref. \cite{martyn} and the interesting discussions.

\end{document}